\documentstyle[twocolumn,psfig]{article}

\makeatletter
\def\section{\@startsection {section}{1}{\z@}{-3.5ex plus -1ex minus 
 -.2ex}{2.3ex plus .2ex}{\large\bf}}
\makeatother

\def\av#1{\langle #1 \rangle}
\def\a{\alpha}
\def\b{\beta}
\def\e{\epsilon}
\def\r{\rho}
\def\s{\sigma}

\begin{document}

\twocolumn[

\begin{center}
{\large \bf Self-consistent perturbational study of insulator-to-metal
transition in Kondo insulators due to strong magnetic field}

{\bf Tetsuro Saso}

{\it Department of Physics, Saitama University, Urawa 338, Japan}
\end{center}

In order to study the effects of strong magnetic field on Kondo
insulators, we calculate magnetization curves and single-particle
excitation spectra of the periodic Anderson model at half-filling
under finite magnetic field by using the self-consistent second-order
perturbation theory combined with the local approximation which
becomes exact in the limit of infinite spatial dimensions.  Without
magnetic field, the system behaves as an insulator with an energy gap,
describing the Kondo insulators. By applying magnetic field to
f-electrons, we found that the energy gap closes and the first order
transition from insulator to metal takes place at a critical field
$H_c$. The magnetization curve shows a jump at $H_c$.  These are
consistent with our previous study in terms of the exact
diagonalization. Relationship to the experiments on YbB$_{12}$ and
some other Kondo insulators is discussed.

\vspace{3mm}
]

\section{Introduction}
The materials called as Kondo insulators show the Kondo behavior at
high temperatures as typical for the heavy fermion metals, but develop
energy gaps at low temperatures.  Examples of such materials are
SmS,\cite{SmS} SmB$_6$,\cite{SmB6} YbB$_{12}$,\cite{YbB12}
TmSe,\cite{TmSe} Ce$_3$Bi$_4$Pt$_3$,\cite{Ce3Bi4Pt3} etc. The energy
gap seems to be formed due to the hybridization of the 4f states with
the conduction bands, but it is considered that the strong correlation
is playing an important role in determining the size of the gap and
the other properties\cite{Aeppli92}. FeSi is also considered to fall
into this family\cite{Aeppli92}.  CeNiSn\cite{CeNiSn} is believed to
have a pseudogap, or even to be a semimetal at lowest
temperatures.\cite{CeNiSn95}

In order to know the characters of the energy gap of the Kondo
insulators, a stimulating experiment was performed on YbB$_{12}$ by
Sugiyama, et al.\cite{Sugiyama88} In their experiment, the application
of a strong magnetic field of about 47T destroys the gap and the
system undergoes a transition from insulator to metal. The
magnetization starts to steeply increase at a critical field. 

It is an interesting theoretical issue to investigate how the strongly
renormalized ground state of the Kondo insulators will be modified and
how the gap is destroyed by the magnetic field. For these purposes, we
previously applied the recently developed techniques to treat the
strongly correlated electron systems in the infinite spatial
dimen\-sions\cite{Metzner89,Georges96}.   In this scheme, the study of
the strongly correlated lattice system is reduced to solving the
impurity problem in an effective medium self-consistently. Therefore,
it is sometimes called as the local-impurity self-consistent
approximation or the dynamical mean field theory.\cite{Georges96}  We
solved the effective impurity problem by the exact diagonalization
(EXD) of the finite system for conduction electrons and calculated the
single-particle excitation spectra to determine the energy
gap.\cite{Saso96,Saso96b,Saso96c} (The paper\cite{Saso96} will be
referred to as I in the following.) We found that the gap decreases
linearly with the magnetic field and vanishes abruptly at a certain
critical field $H_c$, which is of the order of the impurity Kondo
temperature, and that the insulator-to-metal transition takes place.
The magnetization curves exhibit a jump at $H_c$. These features
suggest that the transition is of first order. 

We discussed in I that there are two effects of the magnetic field on
the energy gap in the Kondo insulators. One is the Zeeman shift of the
up- and down-spin bands, which leads to the closing of the gap. The
other is the reduction of the strong renormalization due to
correlation, which results in the increase of the renormalization
factor $z=(1-\partial \Sigma_f(\e)/\partial \e)^{-1}|_{\e=0}$
($\Sigma_f(\e)$ denotes the f-electron self-energy) from $z \ll 1$ to
$z \sim 1$. This gives rise to the {\it increase} of the gap since the
gap $E_g$ is estimated to be of the order of $zV^2/W$, where $V$
denotes the mixing matrix element and $W$ the conduction band width.
We found that at least at half-filling in the symmetric case $z$ does
not change by the field as far as $h<h_c$. Therefore, the energy gap
closes merely by the Zeeman shift of the bands, although an abrupt
transition to metal takes place before the gap completely closes in
this way. This first order transition may be due to the complicated
many-body effects. However, it might be possible that it would be due
to the finite size effect of the conduction electron system.

Recently, Carruzzo and Yu\cite{Carruzzo96} and Tsutsui, et
al.\cite{Tsutsui96} claimed that the gap remains finite even under
strong magnetic field and the field-induced insulator-to-metal
transition does not occur in one dimensional Kondo insulators. Their
conclusions still seem to be indefinite because of their numerical
methods (density matrix renormalization group and exact
diagonalization, respectively). However, it is interesting to study
whether such transition occurs or not in higher dimensions.

In order to clarify these points, we will use in the present paper the
self-consistent second-order perturbation theory
(SCSOPT),\cite{MuellerHartmann89,Schweizer90,Mutou94} and
reinvestigate the character of the field-induced insulator-to-metal
transition in Kondo insulators. As will be explained below, we have
found again that the transition does occur and is of first order, so
that our previous conclusions mentioned above are correct at least
qualitatively. For comparison, a calculation by the iterative
perturbation theory\cite{Georges92} (IPT) was also performed, but we
found that IPT yields quantitatively insufficient results under finite
magnetic field. We will also comment on the possible relationship to
the experiments on YbB$_{12}$ and some other Kondo insulators.

\section{Self-consistent Second-order Perturbation Theory and Iterative 
Perturbation Theory} 
We use the periodic Anderson model (PAM),
\begin{eqnarray}
  H &=& \sum_{ij\s} t_{ij} c_{i\s}^+c_{j\s} + \sum_{i
\s} E_{f\s} f_{i \s}^+ f_{i \s} \nonumber \\
  + & & \hspace{-4mm} \frac{V}{\sqrt{N}} \sum_{i\s} \Bigl( f_{i\s}^+
c_{i\s} + h.c. \Bigr)
  + U \sum_i \delta n_{f i \uparrow} \delta n_{f i \downarrow},
\label{eq:pamha}
\end{eqnarray}
where $\delta n_{f\s} = n_{f\s}-\av{n_{f\s}}$,
$E_{f\s}=E_f+U\av{n_{f-\s}} -\s h$, $h=\mu_B H$ and $H$ denotes the
magnetic field.  $N$ is the number of lattice sites. The other
notations are standard.  We neglect the orbital degeneracy and assume
the symmetric case $E_f=-U/2$ together with the half-filling condition
to express the simplest model to the Kondo insulators.  Since the
g-factors for f- and the conduction electrons ($g_c$ and $g_f$,
respectively) are different, we neglect $g_c$ for simplicity and apply
magnetic field only to f-electrons, although a more general treatment
is possible.\cite{Mutou94} We also assume a paramagnetic ground state.
Possible antiferromagnetic states are regarded as being suppressed by
the introduction of an appropriate frustration.\cite{Rozenberg94}

In SCSOPT\cite{MuellerHartmann89}, the self-energy is calculated as
\begin{eqnarray}
  \Sigma_{f\s}(\e) &=& U^2 \int d\e_1 \int d\e_2 \int d\e_3
\r_\s(\e_1) \r_{-\s}(\e_2) \r_{-\s}(\e_3) \nonumber \\
  & \times &
\frac{f(-\e_1)f(\e_2)f(\e_3)+f(\e_1)f(-\e_2)f(-\e_3)}
     {\e-\e_1+\e_2+\e_3+i\delta} \label{self}
\end{eqnarray}
where $\r_\s(\e)=(-1/\pi)\mbox{Im} G_{f\s}(\e+i\delta)$ is the
f-electron density of states, $f(\e)$ denotes the Fermi function and
\begin{equation}
  G_{f\s}(\e)=\frac{1}{N}\sum_k
\frac{1}{\e-E_{f\s}-\Sigma_{f\s}(\e)-\frac{V^2}{\e-\e_k}}.
\end{equation}
Here $\e_k$ denotes the energy of conduction electrons. These
equations are converted into the following forms:
\begin{eqnarray}
  \Sigma_{f\s}(\e) &=& -iU^2\int_0^\infty\!\!\! dt e^{i\e t} \left[
\b_\s(t) \a_{-\s}(t) \b_{-\s}(t) \right. \nonumber \\
  & & + \left. \a_\s(t) \b_{-\s}(t) \a_{-\s}(t) \right],
\end{eqnarray}
\begin{equation}
  \left. \begin{array}{l} \a_\s(t) \\ \b_\s(t) \end{array} \right\} =
\int_{-\infty}^\infty\!\!\! d\e e^{-i\e t} \r_\s(\e) f(\pm\e),  \label{selfb}
\end{equation}
\begin{equation}
  G_{f\s}(\e)=g_\s(\e) \left[ 1+V^2 g_\s(\e) F(\e-V^2 g_\s(\e))
\right],
\end{equation}
\begin{equation}
g_\s(\e)=(\e-E_{f\s}-\Sigma_{f\s}(\e))^{-1}
\end{equation}
and
\begin{eqnarray}
F(z) &=& \sum_k (z-\e_k)^{-1} \nonumber \\
     &=& (2/W)[z/W-\sqrt{(z/W)^2-1}].
\end{eqnarray}
The last equation was obtained for the semi-circular density of states
of the conduction electrons: $\rho(\e)=(2/\pi
W)[1-\sqrt{1-(\e/W)^2}]$. These equations should be calculated
self-consistently.

In IPT,\cite{Georges92} the lattice problem is converted into solving
an effective impurity problem, so that $\r_\s(\e)$ in eqs.(\ref{self})
and (\ref{selfb}) should be replaced with that for the local Green's
function,
$\tilde{\r}_\s(\e)=(-1/\pi)\mbox{Im}\tilde{G}_{f\s}(\e+i\delta)$,
where
\begin{equation}
  \tilde{G}_{f\s}(\e)=(G_{f\s}(\e)^{-1}+\Sigma_{f\s}(\e))^{-1}.
\end{equation}

\section{Results}
It is rather easy to solve the above equations self-consistently when
$h=0$. We choose $W=1$ as a unit of energy and $V=0.5$. The following
calculations are limited to the absolute zero temperature. In
Fig.1(a), we compare the f-electron density of states $\r(\e)$ by IPT
for $U=2$ with that obtained in I by EXD using the finite
system.\cite{Saso96}
\begin{figure}
\centerline{\psfig{figure=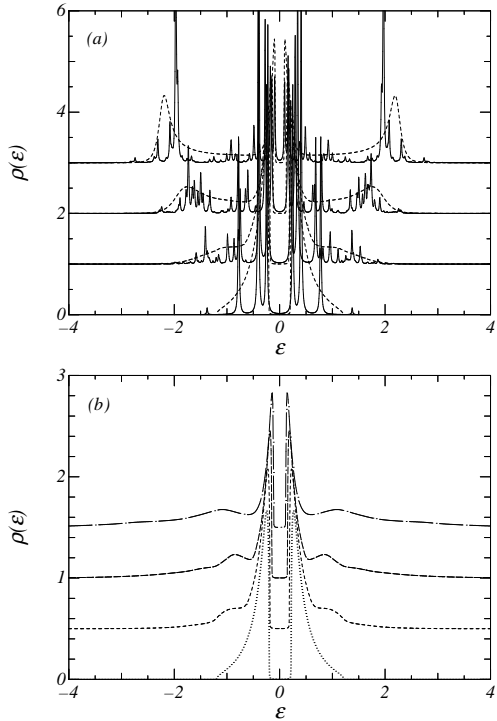,height=12cm}}
\caption{(a) Comparison of the f-electron density of states calculated
by the exact diagonalization of the finite system (full lines)
and by IPT (broken lines) for $U=0,\ 1,\ 2,\ 3$ (from bottom to top). 
(b) The f-electron density of states calculated by SCSOPT for the same
parameters as in (a). Note that the origins of the vertical axis are
shifted.}
\label{fig:1}
\end{figure}
There are rich structures in $\r(\e)$ obtained by EXD. In addition to
the hybridization-split Kondo peaks at small energies, there are peaks
at $\e \sim \pm 0.8$ and $\pm 1.7$ for $U=2$, for instance. The
spectra by SCSOPT are displayed in Fig.1(b). The whole structures are
better reproduced by IPT than SCSOPT. This is in accord with the
results obtained by Mutou and Hirashima\cite{Mutou95} on the
comparison of the quantum Monte Carlo calculation with IPT and SCSOPT.
The peaks at $\e \sim \pm 1.7$ in EXD are the lower and the upper
Hubbard bands, which are pushed out a little beyond $\pm U/2$. In
SCSOPT, there are only small shoulders at these positions, but it is
interesting that there are larger peaks at $\e \sim \pm 0.8$ which
correspond to the peaks in EXD at the same positions.

These differences in the spectra in SCSOPT and IPT come from the
differences in the self-energies. We have plotted the f-electron
self-energies for $U=2$ in Fig.2. It is seen that the absolute values
of the real and the imaginary parts of $\Sigma_f(\e)$ are larger in
ITP than in SCSOPT. This makes the Hubbard bands in IPT be positioned
at the right energies and enhanced sufficiently to reproduce the EXD
results. The fine structures of the spectra in SCSOPT seen above seem
to be originating from the structures in the imaginary part of the
self-energy at the corresponding energy regions.
\begin{figure}
\centerline{\psfig{figure=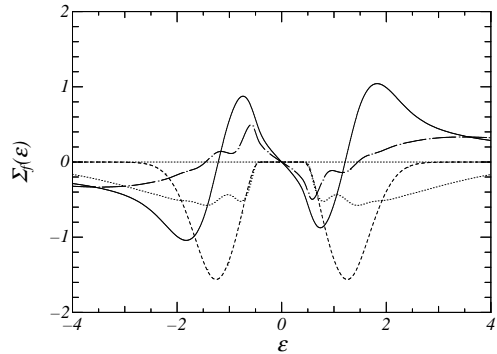,height=6cm}}
\caption{The real and the imaginary parts of the f-electron self-energy 
calculated by IPT (full and broken lines for real and imaginary parts) and
SCSOPT (dash-dotted and dotted lines) for U=2.}
\label{fig:2}
\end{figure}

Despite the better description of the f-electron density of states in
IPT than in SCSOPT, we have found that magnetic properties are better
described by SCSOPT (see below). This is because IPT is not a
conserving approximation.\cite{Mutou94} Therefore, we will use SCSOPT
in the following calculations.

Under finite magnetic field, one also has to determine $n_{f\s}$
self-consistently. We found, however, that the convergence of the
iterative calculation becomes rather poor in the vicinity of the
critical field. To overcome this difficulty, we first assume
appropriate input values $n_{f\s}^0$ for $n_{f\s}$ and solve the above
equations for fixed $E_{f\s}$. Then we obtain $G_{f\s}(\e)$ and
$n_{f\s}$ from them. We repeat this calculation for various
$n_{f\s}^0$'s and plot $m=n_{f\uparrow}-n_{f\downarrow}$ vs.
$m_0=n_{f\uparrow}^0-n_{f\downarrow}^0$ in Fig.3(a).

\begin{figure}
\centerline{\psfig{figure=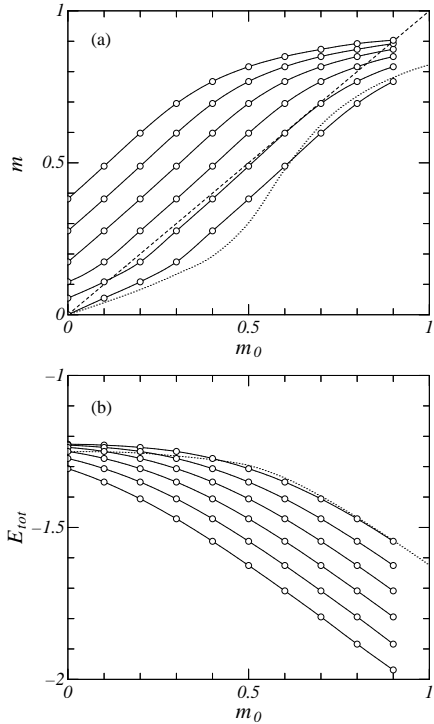,height=12cm}}
\caption{(a) $m$ vs. $m_0$ curves (see the text) obtained by SCSOPT 
for $U=2$ and $h=$0, 0.1, 0.2, 0.3, 0.4, 0.5 from bottom to top. 
(The IPT result
for $h=0$ is displayed by the dotted curve for comparison.) (b) The
total energies for the same set of parameters are plotted from top to
bottom.}
\label{fig:3}
\end{figure}

The points of intersection with the line of the slope of 45 degrees
give the self-consistent solution $m=m_0$. It is clear from the S-like
shape of $m$ vs. $m_0$ curves that the first order transition takes
place at a certain critical field $h_c$, which was found to be
$h_c=0.101$ at $U=2$. We also checked that the total
energy,\cite{Fetter71}
\begin{eqnarray}
 E_{tot}&=&\sum_\s \int d\e f(\e) \left(\frac{1}{\pi}\right) \mbox{Im}
\Bigl[ \e G_{c\s}(\e) \nonumber \\
 & & + \left(\e-\frac{1}{2}\Sigma_{f\s}(\e)\right) G_{f\s}(\e) \Bigr],
\end{eqnarray}
where
\begin{eqnarray}
 G_{c\s}(\e)&=&\sum_k
\frac{1}{\e-\e_k-V^2/(\e-E_{f\s}-\Sigma_{f\s}(\e))} \nonumber \\
 &=&F(\e-V^2g_\s(\e)),
\end{eqnarray}
is lower for larger $m_0$ (Fig.3(b)).

The transition becomes continuous for $U<1.7$ since the S-shape of the
$m$ vs. $m_0$ curves becomes weak. On the other hand, the
ferromagnetic state becomes stable even at $h=0$ for $U>2.6$ because
of a stronger S-shape.

The data in Fig.\ref{fig:3} are spline-interpolated on $m_0$ and $h$,
from which we determine the self-consistent solution $m=m_0$ with a
good accuracy. We plot the magnetization curve in Fig.4.  It shows a
jump at $h_c$ for $1.7 < U < 2.6$, although the size of the jump is
larger than that in EXD. The double steps found by EXD for $U=1$ could
be a reminiscence of a change of the character of the transition from
first to second order.

IPT also yields S-shape curves in the $m$ vs. $m_0$ relations, but the
S-shape is stronger than in SCSOPT (see Fig.3(a), the dotted curve),
giving smaller $h_c$ ($h_c=0.071$ for $U=2$). A fatal drawback in IPT
is that it gives wrong value for the initial slope of the
magnetization curve (i.e. magnetic susceptibility $\chi_f$). We found
$\chi_f=$1.22 and 1.20 in SCSOPT and EXD, respectively, whereas
$\chi_f=0.67$ in IPT for $U=2$. This discrepancy is due to the fact
that IPT is not a conserving approximation\cite{Mutou94} as mentioned
above.
\begin{figure}
\centerline{\psfig{figure=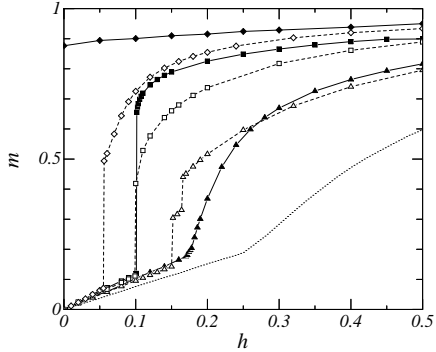,height=6cm}}
\caption{Magnetization curves for $U=$0 (dotted line), 1 (triangles), 
2 (squares), 3 (diamonds) calculated by SCSOPT (filled symbols) and 
EXD (open symbols).}
\label{fig:4}
\end{figure}

The f-electron densities of states under finite magnetic field are
plotted for $U=2$ in Fig.5, from which the energy gaps are obtained
and are plotted in Fig.6 as a function of the field. The gap closes
abruptly between $h=$0.101 and 0.1015. From these results, we consider
that the first order transition obtained in I may not be due to the
finite-size effect.
\begin{figure}
\centerline{\psfig{figure=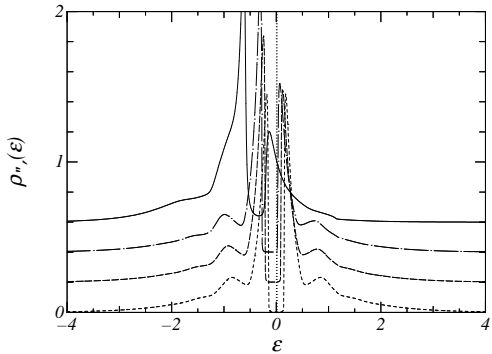,height=6cm}}
\caption{The f-electron density of states for up-spin electrons ($U=2$) 
under the magnetic field $h=$0, 0.05, 0.101 and 0.1015 from bottom to top.}
\label{fig:5}
\end{figure}
\begin{figure}
\centerline{\psfig{figure=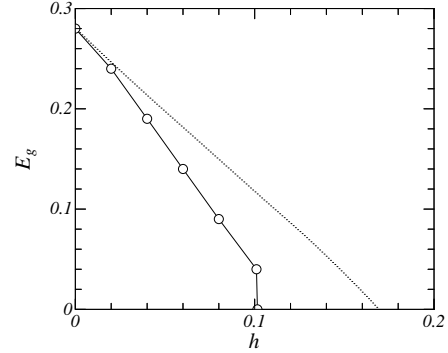,height=6cm}}
\caption{The Energy gap obtained by SCSOPT (circles) as a function of 
$h$ for $U=2$. The dotted line indicates the gap for $U=0$ scaled at $h=0$.}
\label{fig:6}
\end{figure}

Finally, we would like to comment on the behavior of the self-energy
in the magnetic field. Fig.7 shows the field dependence of the real
part of the self-energy. As is seen clearly, Re$\Sigma_{f\s}(\e)$ does
not depend on $h$ when $h<h_c$. This is because the shift of
$\rho_\s(\e)$'s in eq.(\ref{self}) by the magnetic field does not
yield a noticeable change of $\Sigma_{f\s}(\e)$ until the gap is
closed by the sufficient field. Due to this behavior in the
self-energy, the renormalization factor $z$ changes little for
$h<h_c$. Thus the gap closing is mainly caused by the relative Zeeman
shift of up- and down-spin bands. Note that the gap does not close
like $E_g(h)=E_g(0)-2h$ even for $U=0$ since we apply magnetic field
only to f-electrons, so that the shape of $\rho_\s(\e)$ canges with
the field. The energy gap for $h=0$ is given by
$E_g(0)=-W+\sqrt{W^2+4V^2}=0.414$ for $W=1$ and $V=0.5$, whereas $h_c$
is given by $h_c=V^2/W=0.25$, which is larger than $E_g(0)/2$. We have
plotted $E_g(h)$ for $U=0$ in Fig.6 by the dotted line, which is
scaled at $h=0$. The gap by SCSOPT decreases a little faster than that
for $U=0$, which means that the shape of $\rho_\s(\e)$ at $U=2$
changes more than that of $U=0$ by the field despite the small change
of $z$.
\begin{figure}
\centerline{\psfig{figure=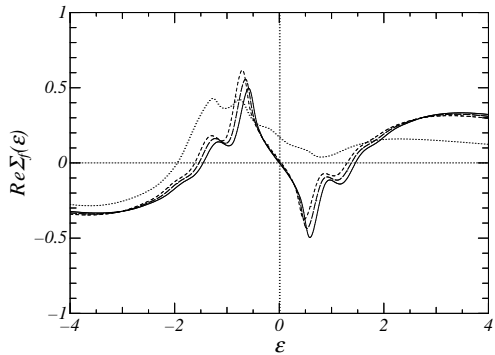,height=6cm}}
\caption{The real parts of the self-energy are plotted for $h=$0 (full 
line), 0.05 (dashed line), 0.101 (broken line) and 0.1015 (dotted line).}
\label{fig:7}
\end{figure}
\section{Discussions}
We have found that the energy gap in Kondo insulators is closed by the
application of a sufficiently strong magnetic field of the order of
the gap, and a transition from insulator to metal does take place in
the limit of infinite spatial dimensions. This is in strong contrast
to the suggestions for the one dimensional Kondo
insulators.\cite{Carruzzo96,Tsutsui96} We found that the transition is
of first order,  and the magnetization shows a jump at the critical
field $h_c$, irrespective of the methods of calculations, although the
first order transition is limited to the range $1.7<U<2.6$ in SCSOPT.
The calculation was only for the electron-hole symmetric case. It is
of much interest to extend the present calculation to more general
situations and to more realistic models with orbital degeneracy. We
started from a paramagnetic ground state, assuming that a frustration
should have destroyed a long range order.\cite{Rozenberg94} It is
necessary, however, to extend the calculation to such an explicit
model which has a paramagnetic ground state.

Despite these issues to be clarified in future studies, we stress that
the magnetic-field-driven insulator-to-metal transition is of great
interest and importance for the study of the characters of the Kondo
insulators since a sufficiently strong magnetic field always tends to
close the gap. On the other hand, the pressure, for instance,
sometimes increases the mixing and the gap, but in some other cases
increases the band width, causing the overlapping of the band and
rendering the system semimetallic. In this sense, the magnetic field
is much simpler and a good tool for the study of the
insulator-to-metal transition in strongly correlated electron systems.

First order transition has not yet been observed in YbB$_{12}$. This
is partly because the energy gap was determined from the temperature
dependence of the resistivity, so that the finite temperature might
have smeared out the transition, or the system is out of range of the
first order transition. More direct experiments at low temperatures,
e.g. measurement of the dynamical conductivity, etc. are desirable.
Lacerda applied the magnetic field of up to 50T to SmB$_6$ and did not
find any indication of a transition to metal.\cite{Lacerda96} This may
be because of the smaller magnetic moment for Sm ($gJ=2.14$ for Ce
whereas $gJ=0.71$ for Sm). Magnetoresistance of CeNiSn is measured to
be positive\cite{CeNiSn95} in contrast to the present calculation. The
structure of the gap and the mechanism of the low temperature
transport in this compound is still not clear. However, the effect of
the magnetic field on the strongly correlated state should be
different if the gap is of V shape. Study of such situation will be
interesting also from the present theoretical point of view.

Recently, a new IPT scheme was proposed\cite{Kajueter96,Kajueter96b},
which can be applied even to the case without electron-hole symmetry.
It may be applicable to a calculation of the properties of the
strongly correlated systems under magnetic field, and may improve the
present calculation, including the cases with larger $U$ and the study
of metamagnetism in the metallic heavy fermion systems.

\section*{Acknowledgements}
The author thank Mr. T. Mutou for a useful comment on the f-electron
density of states. This work is supported by Grant-in-Aid for
Scientific Research on Priority Areas, ``Physics of Strongly
Correlated Metallic Systems",  from the Ministry of Education, Science
and Culture. The computation was done using FACOM VPP500 in the
Supercomputer Center, Institute for Solid State Physics, University of
Tokyo.

\end{document}